# Some Problems of Self-Adjoint Extension in the Schrodinger equation

**Teimuraz Nadareishvili and Anzor Khelashvili**

Institute of High Energy Physics, Iv. Javakhishvili Tbilisi State University, University str. 9,  0109    Tbilisi, Georgia

E-mail: teimuraz.nadareishvili@tsu.ge and anzor.khelashvili@tsu.ge

**Abstract.** The Self-Adjoint Extension in the Schrodinger equation for potentials behaved as an attractive inverse square at the origin is critically reviewed. Original results are also presented.  It is shown that the additional solutions must be retained for definite interval of parameters, which requires performing of Self-Adjoint Extension necessarily. The "Pragmatic approach" is used and some of its consequences are considered for wide class of transitive potentials. The problems of restriction of Self-Adjoint Extension parameter are also discussed. Various relevant applications are presented as well.

## 1.    Introduction

Recently much attention has been devoted to the problems of self –adjoint extension (SAE) for the inverse square ($\frac{1}{r^2}$) behaved potentials in the Schrodinger equation [1]. These problems are interesting not only from academic standpoint. Number of physically significant quantum-mechanical problems manifest in such a behavior. Hamiltonians with inverse square like potentials appear in many systems and they have sufficiently rich physical and mathematical structures. Examples of such systems are:  Valence electron model for hydrogen like atoms in quantum mechanics [2], Coulomb and Hulthen problems in the Klein-Gordon and Dirac equations [3], the theory of black holes [4], conformal quantum mechanics [5], Aharonov-Bohm effect [6], Dirac monopoles [7], quantum Hall effect [8], Calogero model [9] and etc. Mathematical aspects of SAE in differential equations are considered also in [10].
   Detailed consideration of above-mentioned problems puts in doubt the motivations for neglecting of so-called additional (singular) solutions, which are based on mathematical sets of quantum mechanics without invoking of specific physical ideas.
   The aim of this article is to elucidate some vague points, reviewing main papers in this direction. Original results are also presented.

This paper is organized as follows: First, we bring the common reasonings under which these additional solutions are neglected usually. We show that none of them is convinced completely and this problem needs more profound investigation. In Section II, we show that under some circumstances it is necessary to preserve these additional solutions. In Section III the diverse point of view about "falling onto center" is presented. Self –adjoint extension is introduced in Section IV and the last Sections are devoted to various models where this problem takes place.

**2. Statement of Problem**

According to the main hypothesis of the quantum mechanics, eigenvalues of all Hermitian operators are real and observable.(Hermiticity and Self-adjointness are identified in most of textbooks, but this is not correct in general (see,e.g. [11]).

Hermiticity of physical operator puts on some restrictions on its eigenfunctions. For example, from the requirement of hermiticity of Hamiltonian [12] and radial momentum operator $p_r = -i\left\{\dfrac{d}{dr} + \dfrac{1}{r}\right\}$ [13] the following necessary boundary condition results on the radial wave functions at the origin

$$\lim_{r \to 0} rR(r) = \lim_{r \to 0} u(r) = u(0) = 0 \tag{2.1}$$

First of all, we want to pay attention to the significance of this boundary condition. Let's adduce arguments in favour of this condition. W. Pauli in his book [12] remarks, that for hermiticity of Hamiltonian the eigenfunctions, for which $\lim_{r \to 0}(rR) = \lim_{r \to 0} u = A \neq 0$ are inadmissible, while $\int_0^\infty u^* u r^2 dr$ exists. i.e. only normalizability is not sufficient.

The same is underlined in other classical books of quantum mechanics ([14], [15], etc.).

As regards of hermiticity of the radial momentum operator $p_r$, the comment of L.D. Faddeev (in [13], p.336) is important: "Operator $p_r$ is hermitian (symmetric) on functions, that satisfy (2.1), but its extension to self-adjoint one is not possible".

Almost the same ideas are discussed in recent articles. But because of variety of used mathematical requirements, obtained results often differ drastically.

Below we follow to the minimal necessary requirement for hermiticity, i.e. to the boundary condition (2.1) and use the SAE procedure, depending on the behaviour of potential under consideration.

Usually regular potentials are considered in the Schrodinger equation, which obey the following restriction at the origin

$$\lim_{r \to 0} r^2 V(r) = 0 \tag{2.2}$$

In this case the radial wave function behaves as [13-14]

$$\lim_{r \to 0} = C_1 r^l + C_2 r^{-(l+1)} \tag{2.3}$$

where $l$ is orbital momentum. The second term in this expression is singular; it does not satisfy to (2.1) and should be neglected $(C_2 = 0)$.

It is also known, that for singular potentials, that behave like

$$\lim_{r \to 0} r^2 V \to \pm\infty \tag{2.4}$$

"falling onto center" takes place [14-16].

It is interesting to study potentials with intermediate behavior, called "transitive potentials"

$$\lim_{r \to 0} r^2 V \to \pm V_0 \quad (V_0 = const > 0) \tag{2.5}$$

Two signs in the (2.5) correspond to repulsive (+) and attractive (-) potentials, respectively.
For such potentials, the following statement takes place:

**Theorem.** The Schrodinger equation except the standard (non-singular) solutions has also additional solutions for attractive potential, like (2.5). The proof of this theorem is straightforward.

Let us consider radial Schrödinger equation

$$u'' + 2m[E - V(r)]u - \frac{l(l+1)}{r^2} u = 0 \tag{2.6}$$

For the attractive potential in (2.5), this equation for small distances reduces to

$$u'' - \frac{P^2 - 1/4}{r^2} u = 0 \tag{2.7}$$

where

$$P = \sqrt{(l + 1/2)^2 - 2mV_0} > 0 \tag{2.8}$$

Therefore, Eq. (2.7) has following solution

$$\lim_{r \to 0} u = a_{st} r^{1/2+P} + a_{add} r^{1/2-P} = u_{st} + u_{add} \tag{2.9}$$

The standard solutions with $u_{st} \underset{r \to 0}{\sim} r^{1/2+P}$ satisfy to boundary condition (2.1) for arbitrary P. The second term $u_{add} \underset{r \to 0}{\sim} r^{1/2-P}$ is not considered usually, because it is divergent at the origin for $P > 1/2$, but in the interval

$$0 < P < 1/2 \tag{2.10}$$

it also satisfies (2.1) and must be preserved.

For additional states from (2.8) and (2.9) one obtains the condition of the existence of additional states

$$l(l+1) < 2mV_0 \tag{2.11}$$

and if we demand, that P is to be real number (otherwise falling onto center takes place [14-16]) the parameter $V_0$ is restricted by condition

$$2mV_0 < l(l+1) + 1/4 \tag{2.12}$$

The last two inequalities restrict $2mV_0$ in the following interval

$$l(l+1) < 2mV_0 < l(l+1) + 1/4 \tag{2.13}$$

Intervals from the left and from the right sides have no crossing and therefore, if additional solution exists for fixed $V_0$ and for some $l$, then it is absent for another $l$.

Thus from (2.11) we see that in the $l = 0$ states except the standard solutions there are additional solutions as well for arbitrary small $V_0$, while for $l \neq 0$ states the "strong" field is necessary in order to fulfill (2.11).

It should be mentioned, that additional solutions survive such traditional requirement as the normalizability of wave function [16] and the integral from probability density is finite [17]. The stronger restriction on the wave function is considered in monograph [18]. Namely, the matrix elements of kinetic energy operator are required to be finite. To this end, the average value of kinetic energy operator $T = \frac{<\vec{p}^2>}{2m}$ is evaluated by this additional function in $l = 0$ state for a Coulomb potential in the Klein-Gordon equation (This problem after corresponding modifications reduces to the Schrödinger equation with (2.5))

$$<\vec{p}^2> = \int_0^\infty \left(\frac{dR}{dr}\right)^2 r^2 dr \tag{2.14}$$

If we calculate this expression by using $u_{add} \underset{r\to 0}{\sim} r^{1/2-P}$, then it indeed diverges. However, in our opinion this requirement is overestimated. The finiteness of the total energy is sufficient, and indeed, this is the case.

We can demonstrate this result, using generalized virial theorem [19] just for singular potential; It differs from the usual virial theorem and can be written as

$$E = \left\langle V + \frac{1}{2}rV' \right\rangle + \frac{P^2}{m} a_{st} a_{add} \tag{2.15}$$

where $a_{st}$ and $a_{add}$ are given by (2.9). It is evident that for "pure" standard ($a_{add} = 0$) and "pure" additional ($a_{st} = 0$) solutions the usual virial theorem follows from (2.15)

$$E = \left\langle V + \frac{1}{2}rV' \right\rangle \tag{2.16}$$

We see that for our potential (2.5) the total energy is finite. Remember that in the Klein-Gordon equation with Coulomb potential there appears a combination of singular $\left(\frac{1}{r^2}\right)$ and Coulomb-like $\left(\frac{1}{r}\right)$ terms. It is clear from (2.16) that singular parts are cancelled. It is also evident, that the finiteness of total energy follows from explicit calculations as well, without using virial theorem.

Thus, the total energy is finite in case under consideration and the requirement of finiteness of kinetic energy separately is very strong and unjustified.

It can be mentioned that some artificial boundary conditions are also considered in scientific literature. Particularly, restrictions are imposed on wave function and its derivative simultaneously $u(0) = u'(0) = 0$ [20]. It seems to us undesirable from physical point of view, because in this case not only additional but also standard solutions should be forbidden in the range given in (2.10). Author of [21] introduced sterner requirement, namely $\lim_{r\to 0} \frac{u(r)}{\sqrt{r}} = 0$, which is highly artificial, as the Author himself mentions.

There is an interesting remark in the book of R.Newton [22] for (2.5) like potentials. Radial wave function $R = \frac{u}{r}$ has a behavior at small distance

$$R \underset{r\to 0}{\approx} Ar^{-\frac{1}{2}+P} + Br^{-\frac{1}{2}-P} \tag{2.17}$$

Here both terms are singular in the range (2.10). R.Newton pointed out that: "If $P < 1/2$, then the second term is non-regular in the sense that it dominates under the first one. At the same time this non-regular solution is square integrable as well and satisfies to the three – dimensional Schrödinger equation". We think that this argument does not forbid the additional solution.

To summarize all above-mentioned restrictions and comments as well as other artificial ones, we conclude that there is no satisfactory requirement in the framework of quantum mechanics, which avoids this additional solution self-consistently.

Therefore, one has to retain this additional solution and study its consequences.

Self-Adjoint Extension in the Schrodinger equation

## 3. Particle "falling onto center".

First, let us reconsider the problem of particle "falling onto center". It is described in many textbooks and is used in many articles. Most frequently, the book [16] is referenced. In this book, potential of kind (2.5) is regularized near the origin: in the range, $0 \leq r \leq r_0$ this potential is taken as constant and at the end, this regularization is removed $(r_0 \to 0)$. Using this procedure it is argued that the additional solution must be neglected (B = 0 in (2.17)). However, because $u_{add} = Br^{\frac{1}{2}-P}$ is not singular in (2.10) interval, as we think, this regularization and subsequent neglecting is not necessary. We can see it in alternative way.

First let us make some remarks concerning to nodes of wave function. According to well-known theorem for regular potentials (2.2) about the number of nodes for bound states (see, e.g. [13]), the n-th eigenfunction has n-1 nodes (or the ground state eigenfunction does not have nodes). It is easy to show that this theorem remains valid for the attractive potentials like (2.5). Besides that, the second theorem, according to which the number of bound states coincides with the number of nodes of Schrodinger wave function $u(r)$ in $E = 0$ state [13], is also valid for (2.5). Below we consider examples, where these properties are applied.

Let us rewrite equation (2.7) in slightly modified form (in close resemblance to [16])

$$u'' + \frac{\gamma}{r^2} u = 0 \tag{3.1}$$

where

$$\gamma = 2mV_0 - l(l+1) \tag{3.2}$$

This constant is related to above mentioned $P$ as follows

$$P = \sqrt{1/4 - \gamma} \tag{3.3}$$

Let's search the solution of equation (3.1) in the form $u \sim r^{s+1}$. Then we find quadratic equation for s

$$s^2 + s + \gamma = 0 \tag{3.4}$$

with solutions

$$s_1 = -\frac{1}{2} + \sqrt{1/4 - \gamma}; \quad s_2 = -\frac{1}{2} - \sqrt{1/4 - \gamma} \tag{3.5}$$

Consider first the case $0 < \gamma < 1/4$ or $0 < P < 1/2$, when $s_1$ and $s_2$ are real numbers. Thus, the general solution of equation (3.1) should be

$$u = Ar^{s_1+1} + Br^{s_2+1} = Ar^{\frac{1}{2}+P} + Br^{\frac{1}{2}-P} = u_{st} + u_{add} \tag{3.6}$$

Here $u_{add}$ tends to zero at the origin more slowly, than $u_{st}$, but in the interval (2.10) they both have the same properties and must be retained. As one can see in following this causes introduction of self-adjoint extension.

If $\gamma < 0$ or $P > 1/2$, $u_{add}$ does not satisfy (2.1) and one must keep only $u_{st}$.

When $\gamma > 1/4$, or P becomes imaginary number, then $s_1$ and $s_2$ should be mutually conjugated complex numbers

$$s_1 = -\frac{1}{2} + i\sqrt{\gamma - 1/4}; \quad s_2 = s_1^{\bullet} \tag{3.7}$$

In this case the general solution of Eq. (3.1) will be

Self-Adjoint Extension in the Schrodinger equation

$$u \approx A r^{\frac{1}{2}+i\sqrt{\gamma-1/4}} + B r^{\frac{1}{2}-i\sqrt{\gamma-1/4}}$$
$$= A\sqrt{r} \exp[i(\sqrt{\gamma-1/4}\ln r)] + B\sqrt{r}\exp[-i(\sqrt{\gamma-1/4}\ln r)] \qquad (3.8)$$

We see that both solutions oscillate and have same singularity at origin. Taking into account that for for bound states the wave function u must be real, we are forced to require $B^{\bullet} = A$ and therefore

$$u \approx A\sqrt{r}\cos(\sqrt{\gamma-1/4}\ln r + \alpha) \qquad (3.9)$$

where $\alpha$ is an arbitrary constant. Therefore retaining of both solutions causes introduction of "superfluous" parameter $\alpha$, which really is a SAE parameter [23-25]. If we follow the discussion given in [16], we can show that wave function (3.9) corresponds to "falling onto center".

Therefore, it is evident that if we retain $u_{add}$ in $0 < \gamma < 1/4$ domain ($0 < P < 1/2$), the problem of "falling onto center" can be solved without modification (regularization) of potential. It is just the alternate view to this problem.

One can easily confirm that in case $\gamma > 1/4$, the requirement of finiteness of kinetic energy gives the following limitation $\operatorname{Re} s_{1,2} > -\frac{1}{2}$, but now $\operatorname{Re} s_{1,2} = -\frac{1}{2}$. Therefore, in this case both solutions have the same behavior and give infinite kinetic energy. Thus, the argument of authors in Ref. [18] against the additional solution fails.

This problem is well investigated in [23-25]. In particular, both solutions are retained and the self-adjoint (SAE) parameter is introduced. Moreover, for (2.5) like attractive potential the eigenvalue equation for total energy is obtained [23]

$$\gamma_l(E_{nl}) - \gamma_l(E_{0l}) = n\pi, \qquad (n = 0,\pm 1,\pm 2,...) \qquad (3.10)$$

In case of pure inverse square potential, the closed expression for the energy spectrum follows [25]

$$\eta_n = \exp\left[\frac{C - (n+1/2)\pi}{\sqrt{2mV_0 - (l+1/2)^2}}\right] \; ; \; \eta_n = \sqrt{-2mE_n} \; ; \; (n = 0,\pm 1,\pm 2,...) \qquad (3.11)$$

where $\gamma_l$ and C are SAE parameters.

It is natural that retaining the additional solution causes modification of some known results. For example, consider the following potential

$$V = -\frac{V_0}{r^2}, \qquad V_0 > 0$$

in the whole space. There is only one worthy case, namely, $0 < \gamma < 1/4$ or $0 < P < 1/2$. Now the wave function $u$ for $E = 0$ has the form (3.6) in the whole space. It has a single zero, determined by

$$r_0 = \left(-\frac{B}{A}\right)^{1/2P} \qquad (3.12)$$

(It is evident from this relation that constants A and B must have opposite signs in order for $r_0$ to be real number). Therefore, the wave function has only one node and according to above-mentioned theorem we have one bound state only. This result differs from that considered in any textbooks of quantum mechanics (see, e.g. [16]).

We can give very simple physical picture of how the additional solutions arise. For this purpose, let us rewrite the Schrodinger equation near the origin for attractive potential (2.5) in form

Self-Adjoint Extension in the Schrodinger equation

$$u'' + 2m[E - V_{ac}(r)]u = 0 \quad (3.13)$$

where

$$V_{ac} = \frac{P^2 - 1/4}{2mr^2} \quad (3.14)$$

Consider the following possible cases:

i). If $P > 1/2$, then $V_{ac} > 0$ and it is repulsive centrifugal potential and as we saw, one has no additional solutions.

ii). If $0 < P < 1/2$, then $V_{ac} < 0$. Therefore, it becomes attractive and is called as quantum anti-centrifugal potential [25-26]. This potential has $u_{add}$ states, because the condition (2.11) is fulfilled in this case.

iii). If $P^2 < 0$, then $V_{ac}$ becomes strongly attractive and one has "falling onto center".

Therefore, the sign of the potential $V_{ac}$ determines whether we have additional solutions or not.

## 4. Introduction of SAE parameter

As we have shown above for singular attractive potentials like (2.4) and (2.5) in the Schrödinger equation one arbitrary constant like α in (3.9), must be introduced [16, 28-29] for the case

$$2mV_0 > (l + 1/2)^2 \quad (4.1)$$

In mathematical language it means, that the Hamiltonian of this problem is symmetric (Hermitian), but it is not the case for self-adjoint operator. Its defect index [29-30] is (1, 1) and it is necessary to introduce one parameter for SAE, in order for the Hamiltonian to become self-adjoint operator. SAE procedure in mathematics is rather complicated and tedious operation [29-30]. More convenient procedure is an alternative, so-called "pragmatic approach" [31], which gives the same results, as SAE. In particular, it is shown in [31], that if parameter α is the same constant (for fixed $l$), then eigenfunctions of Hamiltonian form a complete orthonormal set and eigenvalues are real, i.e. exactly those properties, which has a self-adjoint Hamiltonian. But (4.1) is a non-physical case, because particle falls onto center, i.e. its energy is unbounded from below.

As to the domain

$$2mV_0 < (l + 1/2)^2 \quad (4.2)$$

one must retain additional $u_{add}$ solutions in the region (2.10). In this case from the Schrödinger equation it follows for arbitrary two levels $E_1$ and $E_2$ and for given $l$, that

$$m(E_2 - E_1)\int_0^\infty u_2 u_1 dr$$
$$= \frac{1}{2}\lim_{r \to 0}\left[u_2(r)\frac{du_1(r)}{dr} - u_1(r)\frac{du_2(r)}{dr}\right] \quad (4.3)$$
$$= P\left(a_1^{st} a_2^{add} - a_2^{st} a_1^{add}\right)$$

The case $P = 0$ must be considered separately, when the general solution of (2.7) behaves as

$$\lim_{r \to 0} u = a_{st} r^{\frac{1}{2}} + a_{add} r^{\frac{1}{2}} \ln r = u_{st} + u_{add} \quad (4.4)$$

Thus, instead of (4.3) one obtains

Self-Adjoint Extension in the Schrodinger equation

$$m(E_2 - E_1)\int_0^\infty u_2 u_1 dr = -\frac{1}{2}\left(a_1^{st} a_2^{add} - a_2^{st} a_1^{add}\right) \qquad (4.5)$$

Hence instead of neglecting of additional solution $(a_{add} = 0)$, as in [25] .for orthogonality in both $P \neq 0$ and $P = 0$ cases one must require

$$a_1^{st} a_2^{add} - a_2^{st} a_1^{add} = 0 \qquad (4.6)$$

Therefore one introduces the SAE $\tau$ parameter

$$\tau \equiv \frac{a_{add}}{a_{st}} \qquad (4.7)$$

So that it is same for all levels (for fixed orbital $l$ momentum, satisfying (2.11)) and is real for bound states.

From (2.9) and (4.7) it is clear that we have three possible cases:

i). $a_{add} = 0$ $(\tau = 0)$. We keep only standard solutions.

ii). $a_{st} = 0$ $(\tau = \pm\infty)$. We keep only additional solutions.

iii). $\tau \neq 0, \pm\infty$. Solutions are neither "pure" standard nor "pure" additional.

We note that in scientific literature [1, 20] for attractive potentials like (2.5), the SAE parameter is always introduced as a necessary attribute. We stress that "pure" $V = -\frac{V_0}{r^2}$ and more generalized (2.5) like attractive potentials give different physical pictures. Therefore in following sections we consider both cases.

## 5. The valence electron model

It is well known that the potential

$$V = -\frac{V_0}{r^2} - \frac{\alpha}{r}; \quad (V_0, \alpha > 0) \qquad (5.1)$$

is used for the description of alkaline metal (Li,Na,K,Rb,Cs) atoms' spectra [2]. At the same time, the similar potential "naturally" arises in the Klein – Gordon equation for the Coulomb interaction, for which SAE will be discussed in future.

The Schrodinger equation for (5.1) in dimensionless variables takes form

$$\left(\frac{d^2}{d\rho^2} - \frac{P^2 - 1/4}{\rho^2} + \frac{\lambda}{\rho} - \frac{1}{4}\right)u = 0 \qquad (5.2)$$

where

$$\rho = \sqrt{-8mE}\,r = ar; \quad \lambda = \frac{2m\alpha}{\sqrt{-8mE}} > 0, \quad E < 0 \qquad (5.3)$$

If we use the notation of [32],

$$u = \rho^{\frac{1}{2}+P} e^{-\frac{\rho}{2}} F(\rho), \qquad (5.4)$$

the equation for confluent hypergeometric functions follows

$$\rho F'' + (2P + 1 - \rho)F' - (1/2 + P - \lambda)F = 0 \qquad (5.5)$$

This equation has four independent solutions, two of which constitute a fundamental system of solutions [33]. They are (in notations of [33]):

Self-Adjoint Extension in the Schrodinger equation

$$\begin{aligned} y_1 &= F(a,b;\rho) \\ y_2 &= \rho^{1-b} F(1+a-b, 2-b; \rho) \\ y_5 &= \Psi(a,b;\rho) \\ y_7 &= e^{\rho} \Psi(b-a, b; -\rho) \end{aligned} \quad (5.6)$$

where

$$a = 1/2 + P - \lambda, \quad b = 1 + 2P \quad (5.7)$$

Only $y_1$ is considered in the scientific articles, as well as in all textbooks (see, e.g. [2, 16, 34]).

Requiring $a = -n$ $(n = 0,1,2,...)$ the standard levels is found. Other solutions $(y_2, y_5, y_7)$ have singular behavior at the origin and usually they are not taken into account. But as was mentioned above, the singularity in case of attractive potentials like (2.5) has the form $r^{\frac{1}{2}-P}$ and in the region $0 < P < 1/2$ other solutions must be considered as well. Therefore, the problem becomes more "rich".

Let us consider a pair $y_1$ and $y_2$. The general solution of (5.5) is

$$u = C_1 \rho^{1/2+P} e^{-\frac{\rho}{2}} F(1/2 + P - \lambda, 1 + 2P; \rho) \\ + C_2 \rho^{1/2-P} e^{-\frac{\rho}{2}} F(1/2 - P - \lambda, 1 - 2P; \rho) \quad (5.8)$$

From the behavior of (5.8) at the origin and from (4.5), we obtain the following expression for SAE $\tau$ parameter

$$\tau = \frac{C_2}{C_1} \frac{1}{(-8mE)^P} \quad (5.9)$$

Note on the other hand that, u must decrease at infinity. From well-known asymptotic properties of confluent hypergeometric function F, we find the following restriction

$$C_1 \frac{\Gamma(1+2P)}{\Gamma(1/2+P-\lambda)} + C_2 \frac{\Gamma(1-2P)}{\Gamma(1/2-P-\lambda)} = 0 \quad (5.10)$$

It gives an equation for eigenvalues in terms of $\tau$ parameter

$$\frac{\Gamma(1/2 - \lambda - P)}{\Gamma(1/2 - \lambda + P)} = -\tau(-8mE)^P \frac{\Gamma(1-2P)}{\Gamma(1+2P)} \quad (5.11)$$

We see that this is very complicated transcendental equation for E, depending on $\tau$ parameter. There are two values of $\tau$, when this equation can be solved analytically:
i) $\tau = 0$. In this case we have only standard levels, which can be found from the condition that $\Gamma(1/2 - \lambda + P)$ has poles

$$1/2 - \lambda + P = -n_r; \quad n_r = 0,1,2... \quad (5.12)$$

ii) $\tau = \pm\infty$. In this case we have only additional levels, obtained from the poles of $\Gamma(1/2 - \lambda - P)$

$$1/2 - \lambda - P = -n_r; \quad n_r = 0,1,2... \quad (5.13)$$

Thus, in these cases i) and ii) one can obtain explicit expression for standard and additional levels

$$E_{st,add} = -\frac{m\alpha^2}{2[1/2 + n_r \pm P]^2} = -\frac{m\alpha^2}{2\left[1/2 + n_r \pm \sqrt{(l+1/2)^2 - 2mV_0}\right]^2} \quad (5.14)$$

where signs (+) or (−) correspond to standard and additional levels, respectively.

Self-Adjoint Extension in the Schrodinger equation

iii) For arbitrary $\tau$ parameter the equation (5.11) is discussed in the Appendix A.

We note that, if we take $V_0 < 0$ in (5.1), then we obtain well-known Kratzer potential [34], but in this case the condition (2.11) is not satisfied. Therefore there are no additional levels for Kratzer potential.

In monographs [2, 34] energy levels for alkaline metal atoms are written in Ballmer's form

$$E_{n'} = -R \frac{1}{n'^2} \tag{5.15}$$

where R is Rydberg constant and $n'$ is the effective principal quantum number

$$n' = n_r + l' + 1 \qquad (n_r = 0,1,2...) \tag{5.16}$$

$l'$ is defined from equation

$$l'(l'+1) = l(l+1) - 8mV_0 \tag{5.17}$$

or

$$l' = -1/2 + P = -1/2 \pm \sqrt{(l+1/2)^2 - 2mV_0} \tag{5.18}$$

Only (+) sign was considered in front of the square root until now.

In [2, 34] $V_0$ was considered to be small and after expansion of this root, the standard levels were derived

$$E_{st} = -R \frac{1}{(n+\Delta_l)^2}; \quad n = n_r + l + 1 \tag{5.19}$$

where

$$\Delta_l \equiv \Delta_l^{st} = -\frac{2mV_0}{2l+1} \tag{5.20}$$

is so - called Rydberg correction (quantum defect) [2,34].

As regards of additional levels, this procedure is invalid, because $V_0$ is bounded from below according to (2.11).

Aapproximate expansion for additional levels is possible only for $l = 0$. We have in this case

$$P = \sqrt{\frac{1}{4} - 2mV_0} \approx \frac{1}{2}(1 - 4mV_0) \tag{5.21}$$

$V_0$ may be arbitrary small, but different from zero, because in this case $P = 1/2$ and we have no levels.

Let us rewrite now the function (5.8) in unit form by using the following relation for the Whittaker function [33]

$$W_{a,b}(x) = e^{-\frac{1}{2}x} x^{\frac{1}{2}+b} \frac{\pi}{\sin \pi (1+2b)} \left[ \frac{F(1/2+b-a, 1+2b; x)}{\Gamma(1/2-a-b)\Gamma(1+2b)} - x^{-2P} \frac{F(1/2-a-b, 1-2b; x)}{\Gamma(1/2+b-a)\Gamma(1-2b)} \right] \tag{5.22}$$

Then from (5.3), (5.8), (5.10) and (5.22) we derive

$$u(r) = C_1 \Gamma(1+2P)\Gamma(1/2-P-\lambda) \frac{\sin \pi (1+2P)}{\pi} W_{\lambda,P}\left(\sqrt{-8mEr}\right) \tag{5.23}$$

Because the Whittaker function $W_{a,b}(x)$ has an exponential damping [35]

Self-Adjoint Extension in the Schrodinger equation

$$W_{a,b}(x) \underset{x \to \infty}{\approx} e^{-\frac{1}{2}x} x^a,$$

(5.24)

it is clear that (5.23) corresponds to a bound state. Moreover, it satisfies the requirement (2.1) in the region (2.10).

Therefore, for $\tau = 0, \pm\infty$ the standard and additional levels are obtained from (5.14) with corresponding wave functions

$$u_{st} = C_1 \rho^{1/2+P} e^{-\frac{\rho}{2}} F(1/2 + P - \lambda, 1 + 2P; \rho) \qquad (5.25)$$

$$u_{add} = C_2 \rho^{1/2-P} e^{-\frac{\rho}{2}} F(1/2 - P - \lambda, 1 - 2P; \rho) \qquad (5.26)$$

For arbitrary $\tau \neq 0, \pm\infty$ the energy can be obtained from the transcendental equation (5.11), while the wave function is given by (5.23).

According to [33] our function (5.23) takes the following form

$$u(r) = C_1 \Gamma(1+2P)\Gamma(1/2 - P - \lambda) \frac{\sin\pi(1+2P)}{\pi} e^{-\frac{\rho}{2}} \rho^{\frac{1}{2}-P} \Psi\left(\frac{1}{2} - \lambda - P, 1 - 2P; \rho\right) \qquad (5.27)$$

where $y_5 = \Psi(a,b,x)$ is one of the above mentioned solutions, (5.6). Its zeros are well-studied [33]: For real $a,b$ (note that in our case $a = \frac{1}{2} - \lambda - P; b = 1 - 2P$ are real numbers) this function has finite numbers of positive roots. However, for the ground state there are three cases where we have no zeros:
1) $a > 0$; 2) $a - b + 1 > 0$; 3) $-1 < a < 0$ and $0 < b < 1$. Only the last case is interesting for us, because $a = \frac{1}{2} - \lambda - P; b = 1 - 2P$ and P is in (2.10). It means

$$-1 < 1/2 - P - \frac{2m\alpha}{\sqrt{-8mE}} < 0 \qquad (5.28)$$

In other words, the ground state energy, which is given by transcendental equation (5.11), must obey this inequality.

Let us now make some comments:
i) One can easily obtain the existences condition on additional levels from (5.19) and (2.11) in diverse form

$$l < \Delta_l < l + 1 \qquad (5.29)$$

If we use data of monograph [34], we obtain that for $l = 0$ states only Li, for $l = 1$ only Ka and for $l = 2$ only Cs satisfy (5.29) (i.e. they have additional solutions and it is necessary to carry out SAE procedure), and Na and Rb have no additional levels.

ii) We have following situation in case of choosing another pairs of solutions of (5.6):
1) ($y_5$ and $y_7$) - do not have levels.
2) ($y_1$ and $y_5$) - give only standard levels (nothing new).
3) ($y_2$ and $y_5$) - give only pure additional levels ($\tau = \pm\infty$), which is unjustified physically, because the standard levels are completely lost.
4) ($y_2$ and $y_7$) - not permissible, because in this case $\tau = 0$ is forbidden and we have no standard levels.

5) ($y_1$ and $y_7$) - not allowed, because in limit $\alpha \to 0$ no levels follow for potential $V = -\dfrac{V_0}{r^2}$, but there exists a single level as we'll see below.

iii) It is interesting to note that Scarf in [36] considered singular potential problem in one dimensional Schrodinger equation

$$V(x) = -\frac{V_0}{x^2} - \frac{\gamma}{xm} \qquad (5.30)$$

So, we have the following correspondence to the potential (5.1):

$$V_0 \to V_0, \qquad \alpha \to \frac{\gamma}{m} \qquad (5.31)$$

He took the general solution as

$$\psi(x) = A\left[u^{(+)}(x) + B_s(\eta) u^{(-)}(x)\right] \qquad (5.32)$$

$$u^{(\pm)}(x) = e^{-\eta x} x^{\frac{1}{2} \pm s} \, {}_1F_1(1/2 \pm s - \gamma/\eta; 1 \pm 2s; 2\eta x) \qquad (5.33)$$

where

$$s = \sqrt{1/4 - U_0}; \quad U_0 = 2mV_0; \quad \eta^2 = -2mE$$

and ${}_1F_1(a, c; z)$ is a confluent hypergeometric function.

These relations coincide with P in (2.8) for $l = 0$ and $\lambda$ in (5.3). Therefore we have

$$\tau_P(E) = B_s(\eta) = -(2\eta)^{-2s} \frac{\Gamma(1 + 2s)\Gamma(1/2 - s - \gamma/\eta)}{\Gamma(1 - 2s)\Gamma(1/2 + s - \gamma/\eta)} \qquad (5.34)$$

According to properties of $\Gamma$ function, the $\tau_P(E)$ is periodic function of E energy with an infinite number of poles and zeroes, which are determined by (5.13) and (5.12), respectively. Moreover the expression for E energy (5.14) coincides with that of (18) from [36]

$$\eta_n^{(\pm)} = \gamma[n + 1/2 \pm s]^{-1}, \quad n = 0,1,2,... \qquad (5.35)$$

However in [36] the SAE procedure is not performed and the additional solutions $\eta_n^{(-)}(B = \infty)$ are neglected "since none of $u_n^{(-)}$ are solutions of Schrodinger equation, when $U_0$ is zero" [36]. But it is incorrect, because in this case we return to the class of regular potentials (2.2).

iv) We note that the problems of additional levels were discussed by other authors as well [37-40]. In particular, in [37] the Klein – Gordon equation is considered with $V = -\dfrac{\alpha}{r}$ Coulomb potential

$$u'' + \left[E^2 - m^2 - \frac{l(l+1)}{r^2} + \frac{2E\alpha}{r} + \frac{\alpha^2}{r^2}\right] u = 0 \qquad (5.30)$$

the author underlines, that there must be levels below the standard levels (called, hydrino eigenstates), which correspond to the expression (5.14) with certain modifications and the (−) sign in front of the root, but these two cases differ from each other. Particularly, it is possible to pass the limit $V_0 \to 0$ in the equation (5.1) and we obtain hydrogen's problem (constants $V_0$ and $\alpha$ are mutually independent), while in (5.30) these constants are not mutually independent and in the limit $\alpha \to 0$ we are faced with the free particle problem, instead of Coulomb's one. Note, that SAE is not performed in the foregoing paper [37].

Moreover this paper is criticized by other authors [41, 42]. Particularly, hydrino states are ignored in [41] for the reason that, for $l = n_r = 0$ from Ballmer's formula does not follow hydrino states in the nonrelativistic limit. But it can be shown that using SAE in the Klein – Gordon equation, the hydrino states correspond to $\tau = \pm\infty$. In [42] it is noticed that hydrino states may be excluded requiring orthogonality, but the detailed study shows that hydrino states must be retained (see Appendix B).

## 6. Inverse square potential.

Let consider another example of SAE in case of inverse square potential

$$V = -\frac{V_0}{r^2}, \quad V_0 > 0 \tag{6.1}$$

It was thought that this potential had no levels out of region of "falling onto center" (See e.g. [13-14, 16]) but in [1, 20, 43, 44] single level was found by complete SAE procedure, while the boundary condition and the range of parameter, like P are questionable. Here we'll show that this potential has exactly a single level, which depends on the SAE parameter $\tau$.

Let's take the Schrodinger equation for (6.1)

$$u'' + \left(-k^2 - \frac{P^2 - 1/4}{r^2}\right) u = 0 \tag{6.2}$$

where P is given by (2.8) and

$$k^2 = -2mE > 0; \quad (E < 0) \tag{6.3}$$

This equation has 3 pairs of independent solutions: $I_P(kr)$ and $I_{-P}(kr)$, $I_P(kr)$ and $e^{i\pi P}K_P(kr)$, $I_{-P}(kr)$ and $e^{i\pi P}K_P(kr)$, where $I_P(kr)$ and $K_P(kr)$ are Bessel and MacDonald modified functions [45]. Consider these possibilities separately.

1) The pair $I_P(kr)$ and $I_{-P}(kr)$:

The general solution of (6.2) is

$$u = \sqrt{kr}\left[AI_P(kr) + BI_{-P}(kr)\right] \tag{6.4}$$

Consider the behaviour of this solution at small and large distances:

a) Small distances

In this case [45]

$$I_P(z) \underset{z\to 0}{\approx} \left(\frac{z}{2}\right)^P \frac{1}{\Gamma(P+1)} \tag{6.5}$$

Then it follows from (6.5) and (6.4) that

$$\lim_{r\to 0} u(r) \approx \sqrt{kr}\left[A\left(\frac{k}{2}\right)^P \frac{r^P}{\Gamma(P+1)} + B\left(\frac{k}{2}\right)^{-P} \frac{r^{-P}}{\Gamma(1-P)}\right] \tag{6.6}$$

From (2.9), (6.6) and the definition (4.7) we obtain of $\tau$ that

$$\tau = \frac{B}{A} 2^{2P} k^{-2P} \frac{\Gamma(1+P)}{\Gamma(1-P)} \tag{6.7}$$

b) Large distances

In this case [45]

$$I_P(z) \underset{z\to\infty}{\approx} \frac{e^z}{\sqrt{2\pi z}} \tag{6.8}$$

and

Self-Adjoint Extension in the Schrodinger equation

$$u(r) \underset{r\to\infty}{\approx} \frac{1}{\sqrt{2\pi}}\{A+B\}e^{kr} \quad (6.9)$$

Therefore, requiring vanishing of $u(r)$ at infinity, we have to take
$$B = -A \quad (6.10)$$
and from (6.7), (6.10) and (6.3) we obtain one real level (for fixed orbital $l$ momentum, satisfying (2.11)),

$$E = -\frac{2}{m}\left[\frac{\Gamma(1+P)}{\Gamma(1-P)}\right]^{\frac{1}{P}}\left[-\frac{1}{\tau}\right]^{\frac{1}{P}} \; ; \quad 0 < P < 1/2 \quad (6.11)$$

Thus, we have derived a level existence of which was pointed out in Sec. II.

Note that exactly this level appears in valence electron model from the eigenvalue equation (5.11) in the limit, $\alpha \to 0$ and because this reason a pair $y_1$ and $y_7$ was excluded in previous section.

Reality of energy in (6.11) restricts $\tau$ parameter to be negative $\tau < 0$. In general $\tau$ is a free parameter but some physical requirements may restrict its magnitude.

Note also that, as it is clear from the derivation of (6.11), this level disappears for $\tau = 0$ and $\tau = \pm\infty$, and for these values scale invariance is restored.

Taking into account a well-known relation [45]

$$K_P(z) = \frac{\pi}{2\sin P\pi}[I_{-P}(z) - I_P(z)] \quad (6.12)$$

we obtain the wave function corresponding to the level (6.11):

$$u = -A\frac{2}{\pi}\sqrt{kr}\sin P\pi \cdot K_P(kr) \quad (6.13)$$

Because of exponential damping

$$K_P(z) \underset{z\to\infty}{\approx} \sqrt{\frac{\pi}{2z}}e^{-z} \quad (6.14)$$

the function (6.13) corresponds to the bound state. It is also known that $K_P(z)$ function has no zeroes for real P $(0 < P < 1/2)$ and therefore (6.13) corresponds to single bound state. Moreover, wave function (6.13) satisfies fundamental condition (2.1).

2) The pair $I_P(kr)$ and $e^{i\pi P}K_P(kr)$;

The general solution of (6.2) is

$$u = \sqrt{kr}\left[AI_P(kr) + Be^{i\pi P}K_P(kr)\right] \quad (6.15)$$

At large distances

$$\lim_{r\to\infty}u(r) \approx \frac{1}{\sqrt{2\pi}}\left(Ae^{kr} + Be^{i\pi P}\pi e^{-kr}\right) \approx A\frac{e^{kr}}{\sqrt{2\pi}} \quad (6.16)$$

Therefore we have no bound states.

The same follows for pair $I_{-P}(kr)$ and $e^{i\pi P}K_P(kr)$. Thus only pair $I_P(kr)$ and $I_{-P}(kr)$ has a single bound state.

We note that:
a) In [44] more general potential is considered:

Self-Adjoint Extension in the Schrodinger equation

$$V(r,\theta,\varphi) = \frac{1}{r^2}\left(\alpha + \beta \cot^2\theta + \nu e^{i\varphi}\right) \tag{6.17}$$

where α,β,ν are arbitrary (in general complex) numbers.

For angular part of wave function the following equation takes place [44]

$$\left[-\frac{1}{\sin\theta}\left(\sin\theta \frac{d}{d\theta}\right) + \frac{\beta\cos^2\theta + m^2}{\sin^2\theta}\right]Y_\eta = \eta Y_\eta \tag{6.18}$$

with solution

$$Y_\eta(\theta,\varphi) = P_\mu^\zeta(\cos\theta) I_{2m}\left(2\sqrt{\nu} e^{\frac{i\varphi}{2}}\right) \tag{6.19}$$

where $\zeta = \sqrt{\beta + m^2}$ and $\mu(\mu+1) = \eta + \beta$, but eigenvalue η is now complex in general. $P_\mu^\zeta$ is the Legendre function and $I_{2m}$ is the modified Bessel function. This time the radial Hamiltonian is non-hermitian, due to the complex potential $\frac{\alpha + \eta}{r^2}$. But if we choose α such that

$$\text{Im}\,\alpha = -\text{Im}\,\eta \tag{6.20}$$

then the potential becomes real $\frac{\alpha + \eta}{r^2} = \frac{\text{Re}\,\alpha + \text{Re}\,\eta}{r^2}$. The effective radial eigenvalue equation

$$\left(-\frac{d^2}{dr^2} - \frac{2}{r}\frac{d}{dr} + \frac{\text{Re}\,\alpha + \text{Re}\,\eta}{r^2}\right) R_{E,\eta}(r) = E R_{E,\eta}(r) \tag{6.21}$$

thus becomes hermitian. Based on his earlier papers [1, 20], the author in [44] obtained a single level, but in our opinion, the allowed region of values of constant $g = l(l+1) + 2m(\text{Re}\,\alpha + \text{Re}\,\eta)$, considered in this paper [44], is questionable

$$-1/2 < g < 3/4 \tag{6.22}$$

Indeed, in the framework of the above described formalism we can obtain single level for potential (6.17) as well, given again by expression (6.11) but now

$$P = \sqrt{1/4 + g} = \sqrt{(l+1/2)^2 - 2m(\text{Re}\,\alpha + \text{Re}\,\eta_R)} \tag{6.23}$$

or using the expression (2.8), we get $V_0 = \text{Re}\,\alpha + \text{Re}\,\eta$. In the region $0 < P < 1/2$ (or $-1/4 < g < 0$) we have a level (6.11), but in the region $1/2 < P < 1$ there are no bound states. Nevertheless, in [44] a level was pointed out in the region $0 < g < 3/4$ (or $1/2 < P < 1$), which may be incorrect, because in the region $1/2 < P < 1$ (or $0 < g < 3/4$) the wave function $u_{add} = a_{add} r^{\frac{1}{2} - P}$ is divergent at the origin and the fundamental boundary condition (2.1) is not satisfied. Thus the statement in Ref. [44] is correct only in the range $0 < P < 1/2$ (or $-1/4 < g < 0$).

Recently the first version of our paper appeared in arXiv: T. Nadareishvili, A. Khelashvili arXiv:0903.0234 (math-ph).v1 2 March 2009 [46], and D,M,Gitman, I.V.Tyutin and B.L.Voronov published their version in arXiv:0903-5277(quant-ph). 30 March,2009 [47 ]. These authors wrote : "A consideration of the Calogero problem requires mathematical accuracy, we disscuss some "paradoxes" inherent in the "naïve" quantum-mechanical treatment». One of such paradoxes is that they obtained a negative energy level in the range $0 < \kappa < 1$, when SAE is performed. (in their notations κ coincides with our P). This conclusion is incorrect, as is based only on normalizability of radial function, but not on boundary condition (2.1), which is valid in the interval $(0 < P < 1/2)$ only. It is the source of the origin of such "paradoxes".

c) In [20] it is noticed that single bound state may be observed experimentally in polar molecules. For example, $H_2S$ and HCl exhibit anomalous electron scattering [48-49], which can be explained only by electron capture. Indeed, for those molecules electron is moving in a point dipole field, and, in this case the problem is reduced to the Schrodinger equation with a potential (6.1). Thus, a level (6.11) obtained theoretically may be observed in those experiments. We note, that while this level was derived in [20], but the boundary conditions $u(0) = u'(0) = 0$ imposed on the wave function remains questionable, as was mentioned above in Sec. II.

d) It was commonly believed, that the potential

$$V = -\frac{V_0}{sh^2 \alpha r} \tag{6.24}$$

has no levels in (4.2) region (see for example problem 4.39 in [50]). In [50] by the arguments of well-known comparison theorem [51-52], which in this case looks like

$$-\frac{V_0}{sh^2 \alpha r} \geq -\frac{V_0}{\alpha^2 r^2} \tag{6.25}$$

it is concluded that the potential (6.24) can not have a level, in the area (4.2), because the potential (6.1) has no levels in this area. But, as we know, there is (6.11) $\tau$ depended one level, therefore the levels for (6.24) are expected. Indeed, in [53] using the Nikiforov-Uvarov method [54], it is shown that (6.24) has infinite number of levels in (4.2).

**7. Problems of restriction of SAE parameter**

As we saw previously the energy E depends on the free parameter $\tau$. Natural question arises: Is it completely free or how can $\tau$ be restricted (if any) based on some physical requirements?

Below we consider several examples for limitation on $\tau$.

*7.1. Problem of level ordering.*

There is well-known theorem [55] about "normal' ordering of levels in the Schrodinger equation according to which the energy of standard levels increases together with increasing number of nodes of wave function for fixed orbital momentum (this theorem follows from well-known Sturm-Luivile comparison theorem for second order ordinary differential equations [56]):

$$E_{st}(n_r + 1, l) > E_{st}(n_r, l) \tag{7.1}$$

From expression (5.14) it is easy to show, that this theorem takes place for additional levels $\tau = \pm \infty$ as well.

It remains to be understood what happens in other points $\tau \neq 0, \pm \infty$. There are two alternatives: If this theorem breaks down for some $\tau$, we can say, that these $\tau$-s must be excluded. On the other hand, however, this fact can be viewed differently. Particularly, we can assume that these values of $\tau$ parameter are also permissible, but the introduction of parameter can change physical picture of the problem.

As regards to another example, in case of singular oscillator

$$V = -\frac{V_0}{r^2} + gr^2 \tag{7.2}$$

it can be shown explicitly, that at $\tau \neq 0, \pm \infty$ the equidistance property $E_{n+1} - E_n = const$ is violated, as well as in the Calogero model [57]. Therefore it is desirable to revise another strong results for the spacing of energy levels. In [58-59] the following theorem for $l = 0$ states is proved

Self-Adjoint Extension in the Schrodinger equation

If
$$Z(r) = \frac{d}{dr} r^5 \frac{d}{dr} \frac{1}{r} \frac{dV}{dr} \geq 0, \quad \frac{dV}{dr} > 0; \quad \forall r; \quad \lim_{r \to 0} r^3 V \to 0 \qquad (7.3)$$
then
$$E(n+1,0) - E(n,0) \geq E(n,0) - E(n-1,0) \qquad (7.4)$$
But for the potential (7.2) we have $Z(r) = 0$ and then from (7.4) the equidistance property follows which as we know, is violated because of SAE.

One can conclude, that the if the fulfillment of equidistance property is required, then only three points $\tau = 0, \pm\infty$ remain.

*7.2. Coulomb repulsion*

Consider the following potential
$$V = -\frac{V_0}{r^2} + \frac{\alpha}{r}; \quad (V_0, \alpha > 0) \qquad (7.3)$$

Let's remember the parameter $\lambda$ given by (5.3), but for our case, the sign in front of $\lambda$ must be reversed in (5.11), i.e. we have
$$\frac{\Gamma(1/2 + \lambda - P)}{\Gamma(1/2 + \lambda + P)} (-8mE)^{-P} = -\tau \frac{\Gamma(1 - 2P)}{\Gamma(1 + 2P)} \qquad (7.4)$$
From this equation it can be seen, that for $\tau = 0$ and $\tau = \pm\infty$ the levels are absent, because $1/2 + \lambda \pm P$ is non-negative integer, unlike the cases (5.11) and (5.12).

For other values of $\tau$ we note that the right-hand side of (7.4) is independent of energy E. Let us examine the behavior of left-hand side for $\varepsilon \to 0$ and for $\varepsilon \to \infty$, where $\varepsilon = -E > 0; E < 0$.

1) $\varepsilon \to 0$ or $\lambda \to \infty$.

Then using the well - known limit [33]
$$\lim_{z \to \infty} \frac{\Gamma(a+z)}{\Gamma(b+z)} \approx z^{a-b} \qquad (7.5)$$
we obtain the left-hand side of (7.4) which tends to the following constant
$$A = \frac{1}{(2m\alpha)^{2P}} > 0 \qquad (7.6)$$

2) $\varepsilon \to \infty$ or $\lambda \to 0$.

In this case, again using the well-known behavior [33]
$$\lim_{x \to 0} \Gamma(a+x) \approx \Gamma(a)[1 + x\Psi(a)]. \qquad (7.7)$$
where $\Psi$ is a logarithmic derivative of Euler's $\Gamma$, we can show, that the left-hand side tends to zero from above.

Therefore, we can conclude that, if the following condition takes place
$$-\tau \frac{\Gamma(1-2P)}{\Gamma(1+2P)} < \frac{1}{(2m\alpha)^P}, \qquad (7.8)$$
there is at least one negative level,

Thus, it seems that the following claim can be made:

If
$$\tau > \tau_0 \qquad (7.9)$$
where
$$\tau_0 = \frac{1}{(2m\alpha)^P} \frac{\Gamma(1+2P)}{\Gamma(1-2P)} \qquad (7.10)$$

then the potential (7.3) has at least one negative level. Indeed, in the limit $\alpha \to 0$ potential (7.3) reduces to potential (6.1) which has one negative level, (6.11). Therefore (7.3) must have at least one level, which is retained in this limit. Therefore, the range of $\tau$, where there are no levels, is unphysical. Thus, $\tau$ is restricted from below by (7.9).

It must be mentioned that, the equation analogous to (7.4) is obtained in [60] for Coulomb interaction in Calogero model. In particular, one negative level in case of Coulomb repulsion is obtained in the framework of full SAE procedure.

The problem becomes more strained in the two-particle Klein-Gordon equation with equal masses in case of Coulomb repulsion with vector and scalar potentials

$$V = \frac{V_0}{r}; \quad S = \frac{S_0}{r}; \quad (V_0 > 0, S_0 > 0) \qquad (7.11)$$

We have the same equation (5.2), but now

$$\lambda = \frac{MV_0/2 + mS_0}{\sqrt{4m^2 - M^2}} > 0; \quad \rho = \sqrt{4m^2 - M^2}\, r; \quad P = \sqrt{(l+1/2)^2 + \frac{S_0^2 - V_0^2}{4}} \qquad (7.12)$$

We must require $4m^2 > M^2$ for bound states.

The eigenvalue equation has the form

$$\frac{\Gamma(1/2 + \lambda - P)}{\Gamma(1/2 + \lambda + P)}(4m^2 - M^2)^{-P} = -\tau\frac{\Gamma(1 - 2P)}{\Gamma(1 + 2P)} \qquad (7.13)$$

From this equation it follows, that there is no bound state levels for $\tau = 0$ and $\tau = \pm\infty$. But if

$$\tau > \tau_0 = -\frac{\Gamma(1 + 2P)}{\Gamma(1 - 2P)}\frac{1}{(MV_0/2 + mS_0)^{2P}}\frac{1}{(4m^2 - M^2)^P} \qquad (7.14)$$

there appears at least one negative level. However, there is one principal difference from previous case (7.3). Here in the limits $V_0 \to 0$, $S_0 \to 0$ the problem reduces to the one of free particles, which has no bound states.

Therefore there are two alternatives for the potential (7.10): we must assume, that there is at least one level for $\tau > \tau_0$, or we must recognize, that the region (7.14) is unphysical and restricts $\tau$ from above

$$\tau < \tau_0 \qquad (7.15)$$

Hence, we conclude that, according to specific physical problems, $\tau$ parameter is somehow constrained, because fixing of $\tau$ is impossible in the framework of mathematical sets of quantum mechanics only.

Finally, the results discussed in this paper suggest that the following **statement** can be verified:
If the Schrodinger equation for regular potential $V(r)$ has no bound states, then the potential

$$W(r) = V(r) - \frac{V_0}{r^2}; \quad (V_0 > 0) \qquad (7.16)$$

does not have bound states for the following values of SAE parameter $\tau = 0, \pm\infty$. But it has at least one bound state in the range of "not - falling onto center"(4.2) for other values of $\tau$.

This statement may be checked in particular cases:
a) $V(r) = 0$

The free particle has no bound state levels. However, as we have shown above, the potential (6.1) has no levels for $\tau = 0, \pm\infty$, but there is a single (one) level (6.11) for another $\tau$-s.

b) $V(r) = \frac{\alpha}{r}; \quad (\alpha > 0)$

Repulsive Coulomb potential has no bound state levels. We have seen above that the potential (7.3) has no levels for $\tau = 0, \pm\infty$, but there is at least one level, if the condition (7.9) is fulfilled.

## 8. Summary

We have shown that for attractive potentials like (2.5) in the Schrodinger equation, it is necessary to keep the so-called additional solutions, because they satisfy all requirements in the range $0 < P < 1/2$ as standard solutions. We described the alternative solution of the problem of "falling onto center", where we have illustrated, that this problem does not require a cut-off regularization, if we keep this additional solution. Retaining the additional solution causes the necessity of SAE procedure. Then, in the framework of "pragmatic approach" or orthogonality requirement, the SAE parameter τ was introduced. In the model of valence electron, the eigenvalue equation depending on τ parameter was derived and investigated. For $\tau = 0$ the well-known form of standard levels followed, but for $\tau = \pm\infty$ the additional levels were obtained. For inverse square potential we found only one negative level, which is absent for $\tau = 0, \pm\infty$. Finally, the free SAE parameter τ was constrained by physical requirements in several examples.

It seems that the performing of SAE procedure is necessary in Schrodinger equation for attractive potentials like (2.5) and for wide class of transitive potentials. This procedure is necessary in various relativistic equations and in scattering problems as well, where the constraint problem of SAE parameter looks more profound.. In this regard, consideration of other dimensions is also interesting. Particularly, the dimension two [26], the excepcional role of which was clarified in paper by K.Kowalsky et al. [63] using apparatus of SAE of symmetric opperators. This and other related problems will be considered in subsequent papers.

**Acknowledgments**

The authors thank Drs. G.Japaridze T.Kereselidze, A.Kvinikhidze M.Nioradze and participants of seminars at Iv. Javakhishvili Tbilisi State University, for many valuable comments and discussions. The designated project has been fulfilled by financial support of the Georgian National Science Foundation (Grant № GNSF/ST07/4-196).

**Appendix A**

Let us investigate transcendental equation (5.11). Note, that within the notations, the left-hand-side of this equation coincides with that of (6.16) of paper [61]. In this paper the 1- dimensional three-body problem with harmonic and inverse square pair potentials is quantized by separating variables in the Schrodinger equation following classical work of Calogero [62], but allowing all possible self-adjoint boundary conditions for angular and radial Hamiltonians. The energy dependence of the left-hand-side of equation (6.16) is studied in detail. Therefore, we can use these results for our equation (5.11). Particularly, let us consider the function

$$F_P(\lambda) = \frac{\Gamma(1/2 - \lambda - P)}{\Gamma(1/2 - \lambda + P)} \tag{A.1}$$

as a function of λ. This function has zeros at

$$\lambda_{n_r}^0 = 1/2 + P + n_r \quad (n_r = 0,1,2...) \tag{A.2}$$

They correspond to $E_{st}$ standard levels of (5.14). $F_P(\lambda)$ becomes $\pm\infty$ at $\lambda_{n_r}^\infty \pm 0$ for

$$\lambda_{n_r}^\infty = 1/2 - P + n_r \quad (n_r = 0,1,2...) \tag{A.3}$$

They correspond to the additional levels $E_{add}$ of (5.14). Using results of Ref. [61], one can show that $F_P(\lambda)$ increases monotonically from

$$F_P(0) = \frac{\Gamma(1/2 - P)}{\Gamma(1/2 + P)} \tag{A.4}$$

to $+\infty$ as $\lambda$ varies from $\lambda = 0$ to $\lambda_{n_r}^\infty$.

Because P is in the interval (2.10), from (A.2) and (A3) the following inequalities

$$\lambda_{n_r}^\infty < \lambda_{n_r}^0 < \lambda_{n_r+1}^\infty; \quad \forall_{n_r} = 0,1,2... \tag{A.5}$$

are fulfilled.

Based on [61] one can show also, that $F_P(\lambda)$ increases monotonically in (A.5) domain. Moreover, this function is negative if $\lambda_{n_r}^\infty < \lambda < \lambda_{n_r}^0$; $\forall_{n_r} = 0,1,2...$ and positive if $\lambda_{n_r}^0 < \lambda < \lambda_{n_r+1}^\infty$; $\forall_{n_r} = 0,1,2...$.

As regards to right- hand - side of (5.10), it may be rewritten in term of $\lambda$ as follows

$$Q_P(\lambda) = -\tau \frac{\Gamma(1-2P)}{\Gamma(1+2P)} (2m\alpha)^{2P} \frac{1}{\lambda^{2P}} \tag{A.6}$$

Therefore, we have the following picture:

For $\tau < 0$, the functions $F_P(\lambda)$ and $Q_P(\lambda)$ intercept each other only once in each $\left[\lambda_{n_r}^\infty, \lambda_{n_r+1}^\infty\right]$ interval. According to definition $\lambda$ (see (5.3)) it means that, we have only one negative level for E. However, in case $\tau > 0$, owing to $F_P(0) > 0$, we have no levels in $\left[0, \lambda_0^\infty\right]$ interval, but in every other interval we have only single negative E level, because of interception of $F_P(\lambda)$ and $Q_P(\lambda)$.

**Appendix B**

As we have noted above, it is thought in [42] that, abandoning of hydrino states may be achieved by requiring orthogonality. In particular, the Schrödinger, one-particle Klein - Gordon and Dirac equations are considered for $V = -\frac{\alpha}{r}$ potential, and it is noted that the respective singular solutions of these equations

$$\lim_{r \to 0} u \approx a_k r^{-l}; k^2 = 2mE \tag{B.1}$$

$$\lim_{r \to 0} u \approx a_k r^{1/2 - P}; k^2 = E^2 - m^2; P = \sqrt{(l + 1/2)^2 - \alpha^2} \tag{B.2}$$

$$\lim_{r \to 0} g, f \approx a_k, b_k r^{-\nu}; k^2 = E^2 - m^2; \nu = \sqrt{(J + 1/2)^2 - \alpha^2} \tag{B.3}$$

do not satisfy to orthogonality conditions, which have the following form for the Schrodinger and Klein – Gordon equations

$$I = \lim_{r \to 0}\left[u_k^*(r) \frac{du_{k'}(r)}{dr} - \frac{du_k^*(r)}{dr} u_{k'}(r)\right] = 0 \tag{B.4}$$

Self-Adjoint Extension in the Schrödinger equation

But the direct calculation using (B.1) yields, that $I \equiv 0$ identically for the Schrodinger equation

$$I = \lim_{r \to 0} r^{-2l-1} (-l)\left(a_k^* a_{k'} - a_k^* a_{k'}\right) \equiv 0 \quad (B.5)$$

Therefore this criterion does not work. On the other hand, the solution (B.1) does not satisfy the fundamental boundary condition (2.1) and by this reason must be neglected.

As regards to the Klein – Gordon equation,

$$I = \lim_{r \to 0} r^{-2P} (1/2 - P)\left(a_k^* a_{k'} - a_k^* a_{k'}\right) \equiv 0 \quad (B.6)$$

But the solution (B.2) again satisfies (2.1) in the (2.10) and therefore the hydrino (additional) states must be retain.

In case of the Dirac equation, again the orthogonality condition has the form

$$I = \lim_{r \to 0}\left(f_k^* g_{k'} - f_{k'} g_k^*\right) = 0 \quad (B.7)$$

In this case, solutions (B.3) do not satisfy (B.7)

$$I = \lim_{r \to 0} r^{-\nu}\left(a_k^* b_{k'} - a_{k'} b_k^*\right) \neq 0$$

Therefore, the result of [42] is correct in this case. It can be easily verified, that the Dirac equation has no hydrino (additional) states.